\documentclass[twocolumn,showpacs,superscriptaddress,amsmath,amssymb,aps,prm]{revtex4-2}
\usepackage{CJK}
\usepackage{lipsum}
\usepackage{times}
\usepackage{graphicx}
\usepackage{subfigure}
\usepackage{dcolumn}
\usepackage{mathrsfs}
\usepackage{amsmath,bm,amsfonts}
\usepackage{latexsym}
\usepackage{textcomp}
\usepackage{pifont}
\usepackage{gensymb}
\usepackage{epstopdf}
\usepackage[perpage]{footmisc} 
\usepackage[breaklinks,colorlinks = true,linkcolor = blue,urlcolor  = blue,citecolor = red,anchor color = green,bookmarks=true]{hyperref}
\usepackage{url} 
\usepackage{hyperref}

\begin{document}

\setlength{\parskip}{0pt}

\title{Theoretical Study of Temperature Dependence of Phonons in Orthorhombic SrZrO$_3$ Perovskite}
\author{P. K. Verma}
\email{pkverma.physics@gmail.com}
\affiliation{Department of Physics, Indian Institute of Science Bangalore, Bangalore 560012, India}

\begin{abstract}
We conduct first-principles theoretical studies to investigate the temperature-dependent phonon properties of orthorhombic SrZrO$_3$ (SZO) perovskite. Our calculations include the quasiharmonic approximation, in which we explored mode Gr\"uneisen parameters, thermal expansion, and frequency shifts for several optical modes. For most modes these shifts exhibit a downward trend with increasing temperature, contributing to the normal behavior of frequencies with temperature. We also studied the temperature dependence of linewidths and lineshifts for several phonon modes within the third-order lattice anharmonic effect. The lineshifts for most modes also exhibit a downward trend with temperature, further supporting the normal temperature-dependent behavior of phonon modes. However, a few optical modes display an upward trend in lineshifts with increasing temperature, which in principle, could lead to anomalous temperature-dependent behavior in their frequencies. Nevertheless, when we incorporate all corrections, including quasiharmonic and third-order anharmonic shifts, to the frequencies obtained within the harmonic approximation, almost all modes exhibit normal behavior with temperature, displaying blue shifts with cooling.

\end{abstract}

\maketitle

\section{\label{sec:intro} Introduction}
Perovskites with the ABO$_3$ crystal structure exhibit a wide range of properties that are both fundamental and technologically significant, such as ferroelectricity \cite{KMRabePF2007}, superconductivity \cite{JGBednorz1988}, and metal-insulator transitions \cite{APRamirezJPCM1997}. SrZrO$_3$ (SZO) is a material of significant technological importance, boasting excellent mechanical, thermal, chemical, and electrical properties such as high dielectric constant, high breakdown strength, low leakage current density, and a wide band gap. Due to these remarkable characteristics, SZO has been extensively studied and is used in several practical applications, including solid oxide fuel cells, hydrogen sensors \cite{YAJIMA1992101,LING2002170}, high-voltage and high-reliability capacitors \cite{KathauserNanoTech2004}, and optical devices \cite{MOREIRA2009293}. Studies of SZO's high-temperature phase transitions have revealed three phases, with the crystal structure changing from orthorhombic $Pnma (Pbnm)$ to orthorhombic $Cmcm$ at 970-1041 K, then to tetragonal $I4/mcm$ at 1020-1130 K, and finally to cubic $Pm\Bar{3}m$ at 1360-1443 K \cite{CarlssonAC1967,AhteeACSB1976,LingyPRB1996,KennedyPRB1999}. To our knowledge, there have only been a few first-principles calculations of the phonon properties in orthorhombic SZO, which revealed that the system does not exhibit any imaginary frequencies \cite{AmisiPRB2012,VALI2008497}.

Vali et al. \cite{VALI2008497} have studied the electronic, vibrational, and dielectric properties of orthorhombic SZO; Evarestov et al. \cite{Evarestov2005R11} examined its structural properties, whereas Liu et al. \cite{LiuSSC2010} investigated its structural, electronic, and optical properties.


Lattice-anharmonic effects play a significant role in determining several physical properties of solids, such as temperature-dependent phonon frequencies, lattice thermal expansion, and phase stability \cite{DCWallace1972}. These effects cause phonons to have a finite lifetime, resulting in materials with finite thermal conductivity. The magnitude of the anharmonicity is highly material-specific, with covalently bonded materials displaying almost negligible anharmonic effects and therefore showing large thermal conductivities \cite{WardPRB2009, LindsayPRB2010}. In contrast, thermoelectric and ferroelectric materials generally exhibit large anharmonicity, as evidenced by their inelastic neutron scattering spectra and ultra-low thermal conductivity values \cite{ZhaoNature2014, ToshiroRMP2014}. In ab initio calculations using the density functional theory (DFT) approach, the lowest-order terms of the phonon self-energies are typically considered \cite{MaradudinPR1962}. For our study on the temperature-dependent phonon properties in SZO perovskite, we have included only the quasiharmonic and third-order anharmonic corrections in our calculations. In the literature, there exist several theoretical investigations of SZO based on first-principles calculations for studying the structural, electronic, elastic, optical, dielectric, vibrational, magnetic, and surface properties. However, most of these studies were focused on the cubic phase of SZO. On the other hand, there are only a few theoretical studies dealing with the lattice dynamical calculations in orthorhombic SZO. To the best of our understanding, this is the first theoretical study dealing with the lattice dynamical calculations including the quasiharmonic and third-order anharmonic corrections in orthorhombic SZO.

The paper is organized as follows: The computational details used in the present work are described in section~\ref{sec:computational}. The section~\ref{sec:anharmonic_details} presents the lattice anharmonic details. Under the results and discussion, the structural properties are presented in section~\ref{subsec:structural}. The zone-center phonon properties within the harmonic approximation are discussed in section~\ref{subsec:vibrational}. The temperature dependence of the lineshifts and linewidths and the frequencies of many optical modes are presented in section~\ref{subsec:anh-contributions}. Here, the mode Gr\"uneisen parameter values are also discussed. The two phonon density of states and the kinematical function are calculated and presented in section~\ref{sec:tphdos_kf}. Finally, the section~\ref{sec:conclusions} presents the conclusions of the paper.

\section{Computational Details}
\label{sec:computational}
In this paper, we used density functional perturbation theory implemented in the QUANTUM ESPRESSO code \cite{HohenbergPR1964, KohnPR1965, BaroniRMP2001, PaoloJPCM2009} to study the lattice dynamical properties of SZO. The interaction between the ionic cores and valence electrons was represented using optimized norm-conserving Vanderbilt pseudopotentials \cite{VanderbiltPRB1990}, and the exchange-correlation energy functional was represented using the local density approximation (LDA) parameterized by the Perdew-Zunger method \cite{CeperleyPRL1980}. We used an energy cutoff of 80 Ry to describe the Kohn-Sham wave functions and carried out k-points summation over the BZ using the Monkhorst-Pack method with a special k-point mesh of $3\times 3\times 3$. For phonon dispersion calculations, we used $\bm q$-grid of $2\times 2\times 2$. We optimized the lattice constants to minimize the total energy, interatomic forces, and unit-cell stresses. The ab-initio calculation of the third-order interatomic force constant matrix elements (IFC3s) within the level of cubic anharmonic approximation is computationally challenging \cite{DebernardiPRL1995, DeinzerPRB2003, GonzePRB1989, DebernardiSSC1994}. To evaluate these IFC3s, we employed the $D3Q$ code \cite{PaulattoPRB2013}, which requires only the linear response of the electronic density to the atomic displacements. We used $\bm q$-grid of $2\times 2\times 2$ to evaluate IFC3s in reciprocal space and then converted them into real space using Fourier transformation. Finally, using these real-space IFC3s and a $\bm q$ mesh of $30\times 30\times 30$, we obtained finer grid-based IFC3s. The same $\bm q$ grid of $30\times 30\times 30$ was used in the quasiharmonic approximation to calculate the temperature-dependent linear thermal expansion coefficient.

\section{Lattice Anharmonic Details}
\label{sec:anharmonic_details}
The temperature dependence of the frequency of $jth$ phonon mode is obtained as
\begin{equation}
\omega_j(T) = \omega_j(0) + \Delta_j^{(qh)}(\omega,T) + \Delta_j^{(3)}(\omega,T) + \dots
\label{eqn:freq_T}
\end{equation}
where $\Delta_j^{(qh)}(\omega,T)$ is the quasiharmonic and $\Delta_j^{(3)}(\omega, T)$ is the cubic anharmonic shifts of frequency of $jth$ phonon mode $\omega_j(0)$. 

The quasiharmonic shift $\Delta_j^{(qh)}$, in Eq.~\ref{eqn:freq_T} accounts for an expansion/contraction of the lattice leading to a change in the harmonic force constants without changing the phonon population and is obtained as \cite{PKVermaPRB2022}
\begin{equation}
\Delta_j^{(qh)}(\omega,T) = \omega_j(0)[e^{-3\gamma_j\int_{0}^{T}\alpha(T')dT'} - 1]
\label{eqn:qh}
\end{equation}
where $\gamma_j$ is Gr\"uneisen parameter of the $jth$ phonon mode, which describes the variations of the vibrational properties of a crystal lattice with respect to its volume, and, as a consequence, the effect that changing temperature has on the size or dynamics of the lattice. 

The linear thermal expansion coefficient in Eq.~\ref{eqn:qh} is calculated as \cite{PKVermaPRB2022}
\begin{equation}
\alpha = \frac{1}{3BV}\sum_{\bm q,j}\left[\gamma_j(\bm q)\frac{[\hbar\omega_j(\bm q)]^2}{k_BT^2}\frac{\exp[\hbar\omega_j(\bm q)/k_BT]}{(\exp[\hbar\omega_j(\bm q)/k_BT]-1)^2}\right]
\label{eqn:alpha}
\end{equation}
where B is the zero-temperature and zero-pressure value of the bulk modulus and $\gamma_j(\bm q)$ is the wavevector-dependent mode Gr\"uneisen parameter of the $jth$ phonon mode.

The third-order shift $\Delta_j^{(3)}(\omega, T)$, in Eq.~\ref{eqn:freq_T} is the real part of the phonon self-energy within the cubic-anharmonic approximation [$\Sigma_j(\omega, T) = \Delta_j^{(3)}(\omega, T) + i \Gamma_j^{(3)}(\omega, T)$, where the linewidth, $\Gamma_j^{(3)}(\omega, T)$, is the imaginary part of self-energy of $jth$ mode] and is obtained by utilizing the Kramers-Kr\"{o}nig relation:
  
\begin{equation}
\Delta_j^{(3)}(\omega_j, T) = -\frac{1}{\pi}\mathcal{P} \int_{-\infty}^{\infty}d\omega'\frac{\Gamma^{(3)}(\omega', T)}{\omega' - \omega_j} 
\label{eqn:anh}
\end{equation}  
where $\mathcal{P}$ stands for the principal value of the integral.

The phonon linewidth in Eq.~\ref{eqn:anh} is obtained as \cite{MaradudinPR1962,ZubarevSPU1960,PascalPRB2006} 
\begin{eqnarray}
\Gamma_j^{(3)}(\omega, T) &=& \frac{18\pi}{\hbar^2}\sum_{\bm{q},j_1,i_2}|V^{(3)}(\bm 0,j;\bm{q},j_1;-\bm{q},j_2)|^2 \nonumber \\
 & &  \times \{[n(\omega_{j_1}(\bm{q}))+ n(\omega_{j_2}(\bm {-q})) + 1] \nonumber \\
 & & \times \delta[\omega - \omega_{j_1}(\bm{q}) - \omega_{j_2}(\bm{-q})] \nonumber \\
& &+ 2[n(\omega_{j_2}(\bm{-q})) -  n(\omega_{j_1}(\bm {q}))] \times \nonumber \\
& & \delta[\omega - \omega_{j_1}(\bm{q}) + \omega_{j_2}(\bm{-q})]  \}
\label{eqn:Gamma}
\end{eqnarray} 

where $V^{(3)}(\bm 0,j;\bm{q},j_1;-\bm{q},j_2)$ is the three phonon coupling constant \cite{PaulattoPRB2013}. The first term in curly brackets on the right-hand side of Eq.~\ref{eqn:Gamma}, is responsible for the non-equilibrium phonon decaying into two low-energy phonons, whereas the second term describes the process in which a non-equilibrium phonon is destroyed together with a thermal phonon and a phonon of higher energy with respect to the initial one is created.

\paragraph*{}    
The factor $ n(\omega_j(\bm q))$ in Eq.~\ref{eqn:Gamma} is the occupation number of the $jth$ phonon mode with wave-vector $\bm q$ and is given by,  
\begin{equation}
n(\omega_j(\bm q)) = \frac{1}{e^{\hbar\omega_j(\bm q)/k_B T} - 1}
\end{equation}   

The Cauchy principal value integral in Eq.~\ref{eqn:anh} and the Dirac delta functions in Eq.~\ref{eqn:Gamma} were evaluated using the Lorentzian broadening function with a broadening parameter of 3 cm$^{-1}$.

\section{Results and Discussion}
\label{sec:results-discussion}

\subsection{Structural Properties}
\label{subsec:structural}

\begin{figure}[htp]
 \centering
 \includegraphics[width=0.3\textwidth]{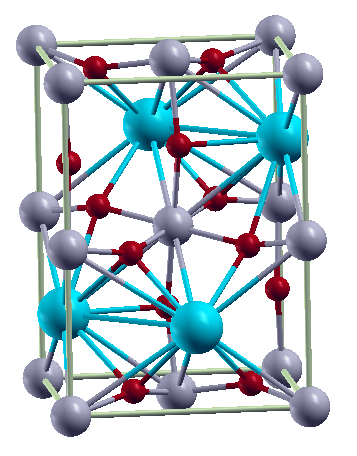}
 \caption{(Color online) Bulk unit cell of orthorhombic SZO. Cyan spheres represent Sr atoms, Zr by grey, and O by red.}
 \label{fig:szo_structure_ortho}
 \end{figure}

The present study focuses on the low-temperature orthorhombic phase of SZO with $Pbnm$ symmetry. The Wyckoff positions of the atoms in this phase are summarized in Table~\ref{table:szo_ortho_wyckoff_pos}. The Zr atom occupies the origin with Wyckoff site 4(a), while the Sr atom is located at $u, 1/2+v, 1/4$ with Wyckoff site 4(c). There are two inequivalent positions for oxygen atoms: O(1) at $u, v, 1/4$ with Wyckoff site 4(c) and O(2) at $1/4 - u, 1/4 + v, w$ with Wyckoff site 8(d). The unit cell of the orthorhombic SZO is illustrated in Figure~\ref{fig:szo_structure_ortho}.

\begin{table}[htp]
\begin{center}
\caption{Atomic displacements $u, v, w$ in orthorhombic SZO according to the space group $Pbnm$ with origin on the Zr atoms. } 
\begin{tabular}{c c c c c}
\hline
 Atom & Wyckoff Position  &  $u$  & $v$   &  $w$  \\
  \hline
  Sr  &  4(c)  &  $u_{Sr}$  &  $\frac{1}{2}+v_{Sr}$  &  $\frac{1}{4}$ \\
  Zr  &  4(a)  &   0   &   0   &   0   \\
  O(1)  &  4(c)  &   $u_{O(1)}$   &  $v_{O(1)}$    &   $\frac{1}{4}$   \\
  O(2)  &  8(d)  &   $\frac{1}{4} - u_{O(2)}$   &  $\frac{1}{4} + v_{O(2)}$   &   $w_{O(2)}$   \\  
\hline
\end{tabular}\label{table:szo_ortho_wyckoff_pos}
\end{center}
\end{table} 

Table~\ref{table:szo_latt_const} presents the calculated lattice constants of SZO and their comparison with the available literature data. The orthorhombic SZO has been the subject of several theoretical studies \cite{LIU2012425,LIU20102032,Evarestov2005R11,VALI2008497,Longo2009FirstPC,LONGO20082191}, and our calculated equilibrium lattice constants are consistent with both theoretical and experimental data \cite{PhysRevB.59.4023,CAVALCANTE20071020,Yamanaka20051496}. Table~\ref{table:szo_atom_pos_cal} shows the calculated atomic positions of orthorhombic SZO, which also agree well with literature values.

\begin{table}[ht]
\centering
\caption{Calculated lattice constants ($a, b$, and $c$) of orthorhombic SZO (in \AA) together with the available theoretical  \cite{LIU2012425,LIU20102032,Evarestov2005R11,VALI2008497,Longo2009FirstPC,LONGO20082191} and experimental data \cite{PhysRevB.59.4023,CAVALCANTE20071020,Yamanaka20051496}.}
\begin{tabular}{c c c c }
\hline
\hline
       &  a      &     b    &   c \\ 
\hline
This work (LDA)  &  5.7025  &  5.7619  &  8.0814 \\
Other calculated results  &   &   &  \\
Theory A \cite{VALI2008497}        &  5.652    &  5.664    &  7.995  \\
Theory B \cite{LIU2012425}  &  5.7077   &  5.7669   & 8.0875  \\
Theory C \cite{LIU20102032}        &  5.8118   &  5.8701   &  8.2426 \\
Theory D \cite{Evarestov2005R11}        &  5.847    &  5.911    &  8.295  \\
Theory E \cite{Longo2009FirstPC}        &  5.78     &  5.79     &  8.15  \\
Theory F \cite{LONGO20082191}        &  5.776    &  5.788    &  8.154  \\
Experimental results   &   &    &   \\
Expt1 \cite{PhysRevB.59.4023}   &  5.7963   &   5.8171   &  8.2048 \\
Expt2 \cite{CAVALCANTE20071020}   &  5.7910   &   5.8108   &  8.1964  \\
Expt3 \cite{Yamanaka20051496}   &  5.816    &   5.813    &  8.225  \\  
\hline
\hline
\end{tabular}
\label{table:szo_latt_const}
\end{table}
%
\begin{table*}[ht]
\centering
\caption{Calculated atomic positions (in \AA) of orthorhombic SZO, compared with the available theoretical  \cite{LIU2012425,VALI2008497,Evarestov2005R11,LiuSSC2010} and experimental data \cite{PhysRevB.59.4023}.}
\begin{tabular}{c c c c c}
\hline
\hline
       &  Sr      &     Zr    &   O(1)   &  O(2) \\ 
\hline
This work (LDA)  &  0.008, 0.535, 0.25  & 0, 0, 0  &  -0.081, -0.022, 0.25  &  0.212, 0.288, 0.043 \\
Other calculated results  &   &   &  & \\
Theory A \cite{VALI2008497}       &  0.007, 0.534, 0.25  &  0, 0, 0  &  -0.107, -0.036, 0.25  &  0.199, 0.301, 0.056 \\
Theory B \cite{LIU2012425}  &  0.008, 0.535, 0.25  &  0, 0, 0  &  0.920, 0.979, 0.25  &  0.213, 0.287, 0.043  \\
Theory C \cite{Evarestov2005R11}       &  0.007, 0.533, 0.25  &  0, 0, 0  &  -0.077, -0.021, 0.25  &  0.213, 0.287, 0.041 \\
Theory D \cite{LiuSSC2010}       &  0.007, 0.519, 0.25  &  0, 0, 0  &  -0.076, -0.0201, 0.25  &  0.2142, 0.2856, 0.0399 \\

Experimental results   &   &    &   & \\
Expt \cite{PhysRevB.59.4023}   &  0.0040, 0.5242, 0.25  &  0, 0, 0  & -0.0687, -0.0133, 0.25  &  0.2154, 0.2837, 0.0363 \\
\hline
\hline
\end{tabular}
\label{table:szo_atom_pos_cal}
\end{table*}

We constructed the pressure-volume curve by optimizing the unit cell at each hydrostatic pressure in the range of 0-30 GPa and fitted it with the BM3 \cite{MurnaghanPNAS1944,BirchPR1947} to obtain the bulk modulus. The resulting value of 172 GPa is very close to the previously calculated value of 170 GPa \cite{LIU2012425} and in reasonably good agreement with the experimentally obtained value of 157 GPa \cite{McKnight_2009,KNIGHT201690}. 

\begin{figure}[ht]
 \centering
 \includegraphics[width=0.45\textwidth]{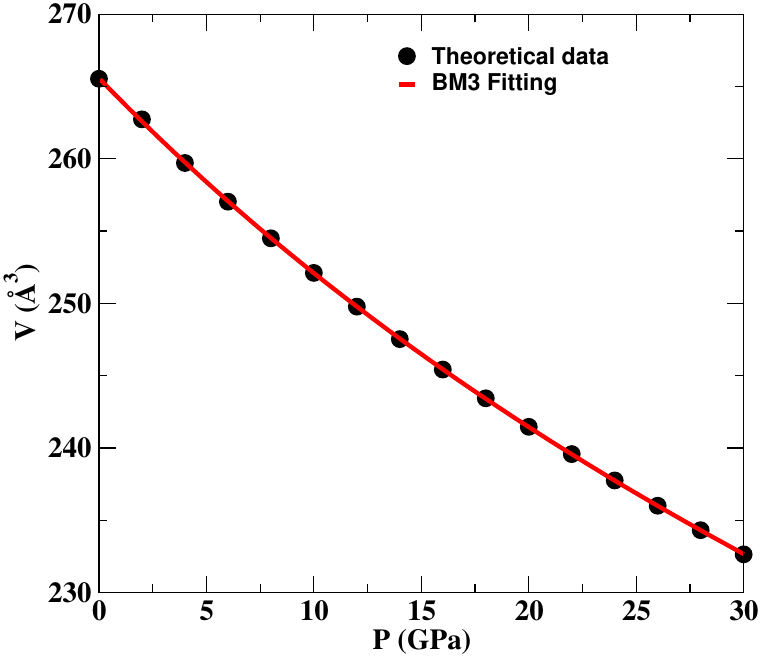}
 \caption{(Color online) Calculated pressure-volume curve of orthorhombic SZO fitted with the BM3. The obtained value of the bulk modulus is 172 GPa.}
 \label{fig:sto_dispersion_phdos}
 \end{figure}

\subsection{Harmonic Lattice Dynamical Properties}
\label{subsec:vibrational}

The orthorhombic SZO has four formula units in its primitive unit cell, belonging to the Pbnm space group, which gives rise to a total of 60 normal modes. The irreducible representations at the BZ center can be expressed as
\begin{equation}
\Gamma = 7 A_g + 5B_{1g} + 7B_{2g} + 5B_{3g} + 8A_u + 10B_{1u} + 8B_{2u} + 10B_{3u} \nonumber
\label{eq:Gamma_irrep}
\end{equation}
Out of these, 24 modes are Raman active ($7 A_g + 5B_{1g} + 7B_{2g} + 5B_{3g}$), 25 modes are IR active ($9B_{1u} + 7B_{2u} + 9B_{3u}$), 3 modes are translational ($B_{1u} + B_{2u} + B_{3u}$), and the remaining 8 modes ($8A_u$) are inactive.

Table~\ref{table:ortho_szo_raman} shows our calculated frequencies of the Raman active modes, along with the literature data. Our mode assignment and calculated frequencies for most modes are consistent with previous theoretical results, especially those based on GGA calculations \cite{AmisiPRB2012,VALI2008497}. However, there are significant differences from previous LDA data by Vali et al. \cite{VALI2008497}, with some modes differing by over 100 cm$^{-1}$. These discrepancies may be due to differences in the calculation parameters used in the two studies. Our calculated frequencies for most modes are in good agreement with experimental values \cite{OKamishima_1999}, although previous experiments did not observe all $B_{1g}, B_{2g}$, and $B_{3g}$ modes, and the three modes with calculated frequencies of 730, 770, and 796 cm$^{-1}$ were not observed due to experimental limitations on frequencies below 600 cm$^{-1}$.

\begin{table}[ht]
\centering
\caption{Calculated frequencies (cm$^{-1}$) of the Raman active modes of the orthorhombic SZO together with the available theoretical and experimental data.}
\begin{tabular}{c c c c c}
\hline
 Mode & This work (LDA) &  GGA \cite{AmisiPRB2012} & LDA\cite{VALI2008497}  &  Expt. \cite{OKamishima_1999} \\
 \hline
 $A_{g}$  &  103  & 94  & 135  &  96 \\
 $A_{g}$  &  112  &  110 & 214  & 107  \\
 $B_{2g}$  & 125   &  119 & 161  &  117 \\
 $B_{1g}$  & 132   & 128  & 181  &  133 \\
 $B_{2g}$  & 147   & 136  & 202  &   \\
 $B_{3g}$  & 151   & 140  & 192  & 138  \\
 $B_{2g}$  & 158   & 145  & 314  & 146  \\
 $B_{3g}$  & 159   & 155  & 333  &   \\
 $A_{g}$  &  187  &  174 & 287  &  169 \\
 $A_{g}$  &  264  & 254  & 356  &  242 \\
 $B_{1g}$  & 285   & 281  & 534  &   \\
 $A_{g}$  &  299  & 278  & 261  &  278 \\
 $B_{3g}$  & 327   & 310  & 522  & 315  \\
 $B_{2g}$  & 331   & 319  & 413  &   \\
 $B_{2g}$  & 385   & 376  & 650  & 392  \\
 $B_{1g}$  & 401   & 390  & 526  &   \\
 $A_{g}$  &  409  & 398  & 632  & 413  \\
 $B_{2g}$  & 439   & 426  & 672  & 441  \\
 $B_{3g}$  & 564   & 528  & 649  &   \\
 $B_{1g}$  & 566   & 530  & 593  & 547  \\
 $A_{g}$  & 571   & 537  & 672  & 556  \\
 $B_{3g}$  & 730   & 708  & 760  &   \\
 $B_{2g}$  & 770   & 750  & 751  &   \\
 $B_{1g}$  & 796   & 777  & 709  &   \\ 
\hline
\end{tabular}
\label{table:ortho_szo_raman}
\end{table}

Table~\ref{table:ortho_szo_ir} presents the frequencies of transverse optic IR active phonon modes. As no IR measurements have been reported in the literature for this crystal structure, we compare our results with previous studies based on GGA \cite{AmisiPRB2012} and LDA \cite{VALI2008497} methods. Our calculated frequencies are in excellent agreement with the GGA results reported by Amisi et al. \cite{AmisiPRB2012}. However, we observe significant differences from the LDA data reported by Vali et al. \cite{VALI2008497}, with some mode frequencies differing by over 100 cm$^{-1}$.

Table~\ref{table:ortho_szo_silent} reports the frequencies of optically inactive phonon modes in orthorhombic SZO. Here, we only had access to LDA data from Vali et al. \cite{VALI2008497} for comparison, and we again observe significant discrepancies in the frequency values.

\begin{table}[ht]
\centering
\caption{Calculated frequencies (cm$^{-1}$) of IR modes of the orthorhombic SZO compared with other theoretical data.}
\begin{tabular}{c c c c }
\hline
 Mode & This work (LDA) & GGA \cite{AmisiPRB2012} & LDA \cite{VALI2008497} \\
 \hline
 $B_{1u}$  &  100  & 94    & 54 \\
 $B_{3u}$  &  106  & 102   & 108 \\
 $B_{2u}$  &  113  & 111   & 141  \\
 $B_{2u}$  &  137  & 129   & 186 \\
 $B_{3u}$  &  153  & 140   & 205 \\
 $B_{1u}$  &  155  & 149   & 204 \\
 $B_{3u}$  &  193  & 184   & 230  \\
 $B_{1u}$  &  204  & 188   & 270  \\
 $B_{3u}$  &  209  & 192   & 236  \\
 $B_{2u}$  &  211  & 199   & 315  \\
 $B_{1u}$  &  213  & 203   & 298  \\
 $B_{2u}$  &  244  & 225   & 311  \\
 $B_{3u}$  &  256  & 240   & 352  \\
 $B_{1u}$  &  260  & 252   & 339  \\
 $B_{3u}$  &  279  & 265   & 373  \\
 $B_{1u}$  &  329  & 308   & 372  \\
 $B_{2u}$  &  333  & 317   & 416  \\
 $B_{1u}$  &  339  & 320   & 450  \\
 $B_{1u}$  &  364  & 356   & 637  \\
 $B_{3u}$  &  386  & 374   & 457  \\
 $B_{3u}$  &  462  & 448   & 649  \\
 $B_{2u}$  &  492  & 459   & 456  \\
 $B_{3u}$  &  505  & 474   & 666  \\
 $B_{2u}$  &  512  & 481   & 621  \\
 $B_{1u}$  &  521  & 491   & 668  \\
 \hline
\end{tabular}
\label{table:ortho_szo_ir}
\end{table}

\begin{table}[ht]
\centering
\caption{Calculated frequencies (cm$^{-1}$) of silent modes of the orthorhombic SZO compared with other theoretical data.}
\begin{tabular}{c c c}
\hline
 Mode & This work (LDA) & LDA \cite{VALI2008497} \\
 \hline
 $A{u}$  &  98   & 82   \\
 $A{u}$  &  131  & 149   \\
 $A{u}$  &  148  & 161   \\
 $A{u}$  &  195  & 259  \\
 $A{u}$  &  245  & 297   \\
 $A{u}$  &  328  & 420   \\
 $A{u}$  &  487  & 434  \\
 $A{u}$  &  509  & 636   \\ 
 \hline
\end{tabular}
\label{table:ortho_szo_silent}
\end{table}

\begin{figure}[ht]
  \subfigure[$B_{1u}$(100 cm$^{-1}$).]{\includegraphics[width=.26\linewidth]{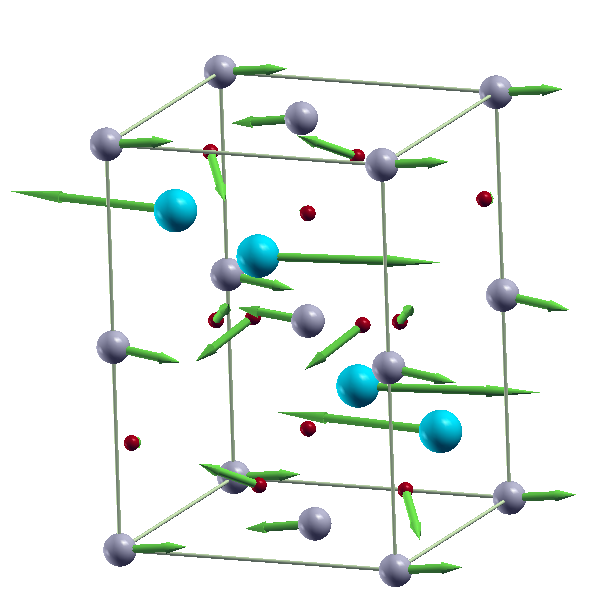}}
  \subfigure[$A_g$(103 cm$^{-1}$).]{\includegraphics[width=.24\linewidth]{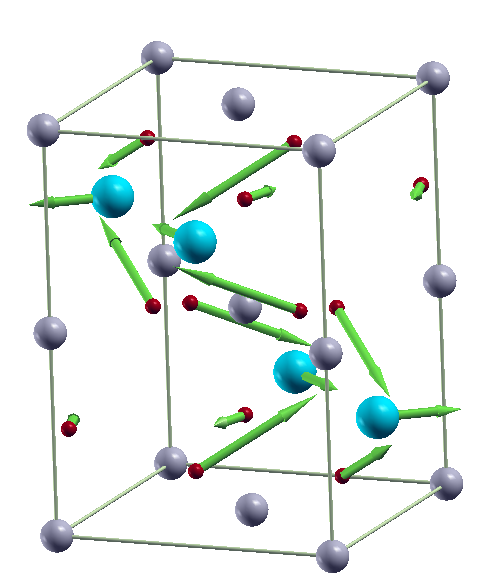}}
  \subfigure[$B_{3u}$(106 cm$^{-1}$).]{\includegraphics[width=.24\linewidth]{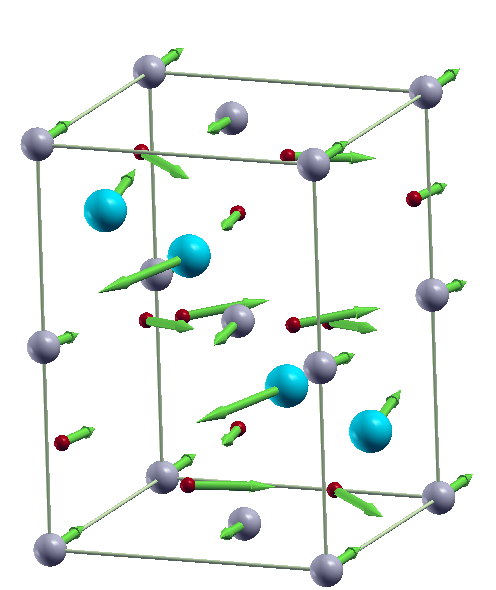}}  \\
  \subfigure[$B_{2u}$(113 cm$^{-1}$).]{\includegraphics[width=.26\linewidth]{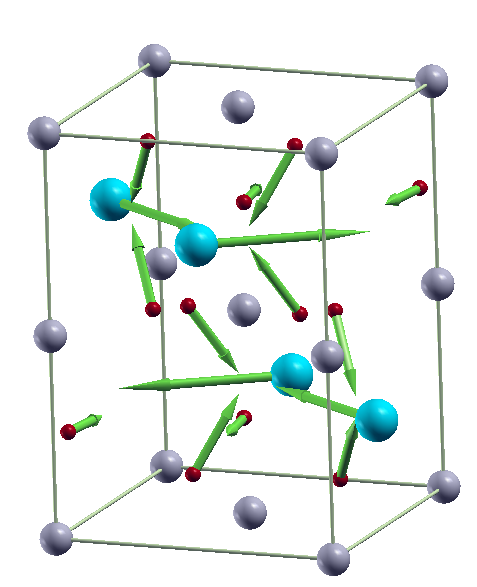}}
  \subfigure[$B_{2g}$(125 cm$^{-1}$).]{\includegraphics[width=.26\linewidth]{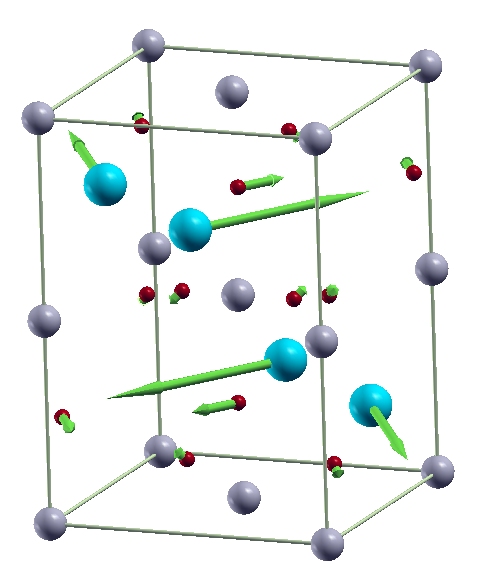}}
  \subfigure[$B_{2g}$(439 cm$^{-1}$).]{\includegraphics[width=.26\linewidth]{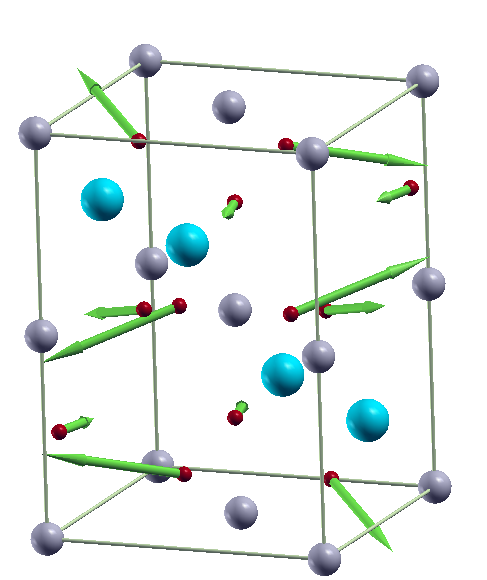}} \\
  \subfigure[$B_{3u}$(505 cm$^{-1}$).]{\includegraphics[width=.26\linewidth]{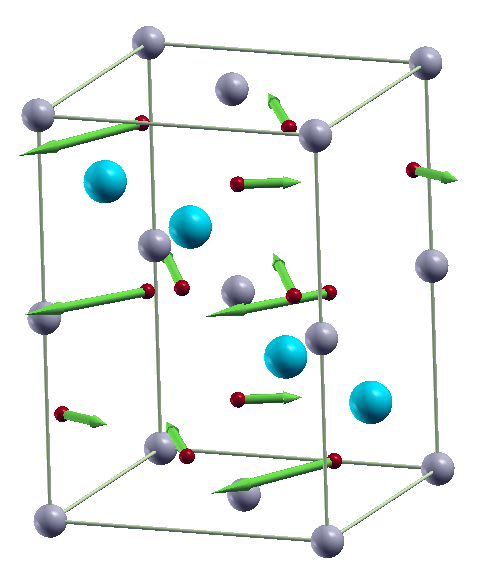}}
  \subfigure[$B_{2u}$(512 cm$^{-1}$).]{\includegraphics[width=.26\linewidth]{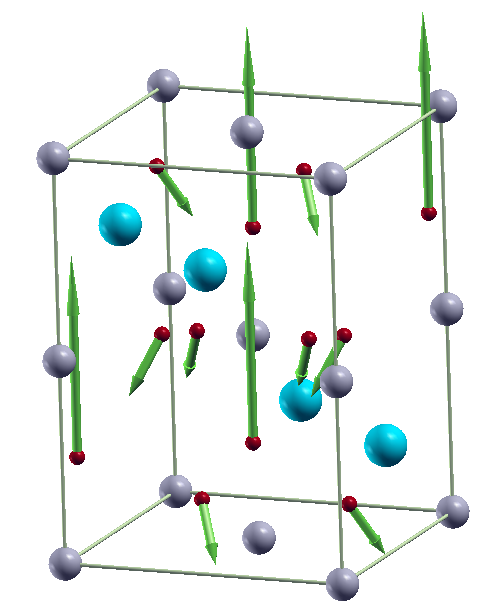}}
  \subfigure[$B_{1u}$(521 cm$^{-1}$).]{\includegraphics[width=.26\linewidth]{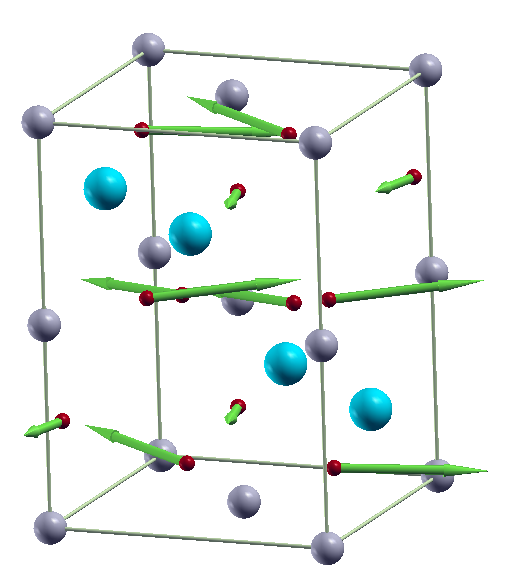}}  \\
  \subfigure[$B_{3g}$(730 cm$^{-1}$).]{\includegraphics[width=.26\linewidth]{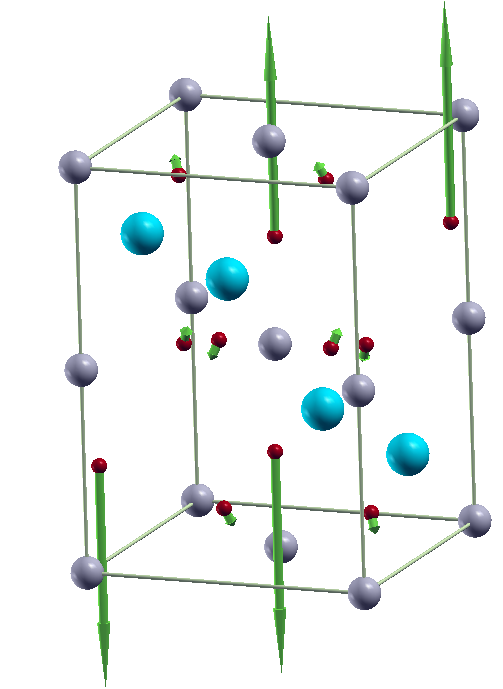}}
  \subfigure[$B_{2g}$(770 cm$^{-1}$).]{\includegraphics[width=.26\linewidth]{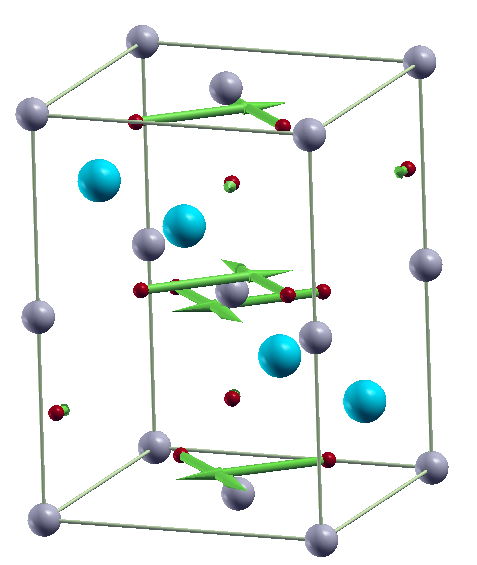}}
  \subfigure[$B_{1g}$(796 cm$^{-1}$).]{\includegraphics[width=.26\linewidth]{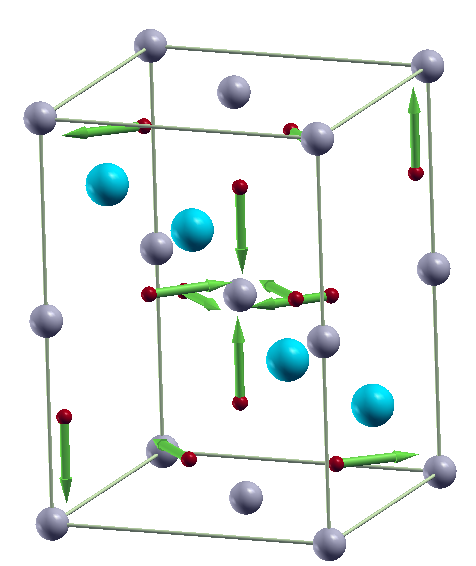}}
\caption{(Color online) Displacement patterns of the optical modes of orthorhombic SZO. The  Zr atoms are represented by grey spheres, Sr atoms by cyan spheres, and O atoms by red spheres. The bonds are removed for clarity.}
\label{fig:szo_displacement_vectors}
\end{figure}

Figure~\ref{fig:szo_displacement_vectors} shows the displacement patterns of twelve optical modes in orthorhombic SZO, with the atomic bonds removed for clarity. The first five low-frequency modes involve the vibrations of both cations and anions, while the remaining high-frequency modes involve only anion vibrations.

\subsection{Temperature Dependent Contributions}
\label{subsec:anh-contributions}

We investigate the temperature dependence of phonon properties of orthorhombic SZO in the range of 0-300 K. The mode Gr\"uneisen parameters (required to evaluate the temperature dependence of the quasiharmonic shifts in Eq.~\ref{eqn:qh}), is calculated as $\gamma_j = -\partial\ln\omega_j/\partial\ln V$, where $\omega_j$ is the frequency of $jth$ mode and $V$ is the cell volume. The zone center phonon calculations were performed at several values of the cell volumes to construct the $\omega_j(V)$ versus $V$ curve, which was then fitted with the cubic spline polynomials to obtain the smooth curve, from which the derivatives of the frequency with volume were evaluated. Thus obtained the mode Gr\"uneisen parameters of Raman active modes are shown in Table~\ref{table:szo_gamma_delta3_raman}. Notably, the mode Gr\"uneisen parameters for modes $A{g} (103 \text{cm}^{-1})$, $B{3g} (151 \text{cm}^{-1})$, $B_{2g} (158 \text{cm}^{-1})$, and $A_{g} (264 \text{cm}^{-1})$ have negative values of -1.29, -1.0, -0.96, and -0.31, respectively, which arise due to the softening of the frequency of these modes with the application of pressures. The mode Gr\"uneisen parameters for the IR and optically inactive modes are shown in the Tables~\ref{table:szo_gamma_delta3_ir} and ~\ref{table:szo_gamma_delta3_silent}, respectively. Notably, certain low-frequency modes have negative values for their mode Gr\"uneisen parameters, which again arise due to the softening of the frequency of those modes with pressure. It is worth noting that, to our knowledge, there is no prior literature data available for comparison with our results.

Next, the linewidths, lineshifts, and $d\Gamma^{(3)}_j = \Gamma_j^{(3)}(300 \text{K}) - \Gamma_j^{(3)}(0 \text{K})$ and $d\Delta_j^{(3)} = \Delta_j^{(3)}(300 \text{K}) - \Delta_j^{(3)}(0 \text{K})$ for the Raman active modes are also shown in Table~\ref{table:szo_gamma_delta3_raman}. We notice that the linewidths at zero Kelvin, $\Gamma_j^{(3)}$ (0 K), are significantly smaller for low-frequency modes compared to high-frequency ones. However, the change in the linewidths of low-frequency modes from 0 K to 300 K, $d\Gamma^{(3)}_j$, is comparable to that of high-frequency modes. Additionally, the linewidths and lineshifts are non-zero at zero Kelvin due to zero-point corrections in energies~\cite{PKVermaPRB2022}. We note that the change in lineshifts from 0 K to 300 K, $d\Delta^{(3)}_j$, is slightly larger for low-frequency modes than high-frequency ones. Of further interest are the positive values of $d\Delta_j^{(3)}$ observed for some high-frequency modes, namely $B_{2g} (439 \text{cm}^{-1})$, $B_{3g} (730 \text{cm}^{-1})$, $B_{2g} (770 \text{cm}^{-1})$, and $B_{1g} (796 \text{cm}^{-1})$, with values of 0.21, 2.33, 0.97, and 0.48 cm$^{-1}$, respectively. The positive lineshifts, in principle, could lead to the blue shift in frequencies with increasing temperature, formally known as the anomalous temperature dependence of the modes. However, the positive shifts in the present case are quite small and the overall temperature dependence of the frequency of any mode also depends on the quasiharmonic shifts and the higher-order anharmonic shifts. The higher-order terms, beyond the third-order, are not considered in the present study. Furthermore, the linewidths and lineshifts for the IR and optically inactive modes are shown in Tables~\ref{table:szo_gamma_delta3_ir} and ~\ref{table:szo_gamma_delta3_silent}, respectively.

\begin{table*}[htp]
\centering
\caption{SZO data of Raman frequencies at $T = 0 $ K, mode Gr\"uneisen parameter ($\gamma_j$),  linewidths ($\Gamma_j^{(3)}(T)$) and lineshifts ($\Delta_j^{(3)}(T)$) calculated at $T = 0$ and $T = 300$ K and their differences, i.e. $d\Gamma^{(3)}_j = \Gamma_j^{(3)}(300 \text{K}) - \Gamma_j^{(3)}(0 \text{K})$ and $d\Delta^{(3)}_j = \Delta_j^{(3)}(300 \text{K}) - \Delta_j^{(3)}(0 \text{K})$. Frequencies, linewidths, and lineshifts are given in units of cm$^{-1}$.}
\begin{tabular}{c c c c c c c c c}
\hline
Mode & $\omega$ & $\gamma_j$ &$\Gamma_j^{(3)}$ (0 K) & $\Gamma_j^{(3)}$ (300 K)  & $d\Gamma^{(3)}_j$  & $\Delta_j^{(3)}$ (0 K)  & $\Delta_j^{(3)}$ (300 K)  &  $d\Delta^{(3)}_j$ \\
\hline
$A_{g}$   & 103   & -1.29  &  0.02   & 3.66   & 3.65   &  -1.68   &  -8.77   & -7.09   \\
$A_{g}$   & 112   &  2.30  &  0.18   & 4.55   & 4.36   & -1.92   & -8.69   & -6.77  \\
$B_{2g}$  & 125   &  1.19  &  0.04   & 1.15   & 1.12   & -0.78   & -3.09   & -2.31  \\
$B_{1g}$  & 132   &  2.72  &  0.08   & 1.81   & 1.73   & -0.87   & -3.47   & -2.60  \\
$B_{2g}$  & 147   &  0.62  &  0.06   & 1.23   & 1.17   & -0.70   & -2.74   & -2.03  \\
$B_{3g}$  & 151   & -1.00  &  0.25   & 4.03   & 3.77   & -1.93   & -6.71   & -4.78  \\
$B_{2g}$  & 158   & -0.96  & 0.15    & 1.27   & 1.11   & -0.45   & -1.46   & -1.01  \\
$B_{3g}$  & 159   &  0.62  & 0.23    & 3.38   & 3.15   & -1.05   & -4.67   & -3.62  \\
$A_{g}$   &  187  &  0.33  & 0.23    & 2.83   & 2.60   & -1.48   & -4.48   & -3.00  \\
$A_{g}$   &  264  & -0.31  & 0.44    & 3.33   & 2.89   & -2.09   & -4.57   & -2.48  \\
$B_{1g}$  &  285  &  0.15  & 0.70    & 4.07   & 3.37   & -2.17   & -4.05   & -1.88  \\
$A_{g}$   &  299  &  0.15  & 1.41    & 6.22   & 4.80   & -2.13   & -2.30   & -0.17  \\
$B_{3g}$  &  327  &  0.82  &  0.86   & 5.34   & 4.47   & -2.12   & -3.83   & -1.72  \\
$B_{2g}$  &  331  &  0.53  & 0.67    & 3.19   & 2.53   & -2.20   & -4.25   & -2.04  \\
$B_{2g}$  &  385  &  0.34  & 1.02    & 4.82   & 3.80   & -2.52   & -4.31   & -1.79  \\
$B_{1g}$  &  401  &  0.50  & 0.98    & 3.37   & 2.39   & -1.36   & -1.48   & -0.13  \\
$A_{g}$   &  409  &  0.77  & 0.85    & 3.54   & 2.69   & -1.45   & -1.56   & -0.11  \\
$B_{2g}$  &  439  &  0.67  &  1.45   & 4.38   & 2.93   & -1.30   & -1.08   & 0.21  \\
$B_{3g}$  &  564  &  2.78  & 3.19    & 9.48   & 6.29   & -4.74   & -7.31   & -2.57  \\
$B_{1g}$  &  566  &  2.52  & 3.61    & 9.78   & 6.17   & -4.57   & -5.99   & -1.42  \\
$A_{g}$   &  571  &  2.74  & 3.23    & 8.70   & 5.47   & -4.06   & -5.50   & -1.44  \\
$B_{3g}$  &  730  &  1.85  & 5.05    & 10.16   & 5.12   & -3.36   & -1.04   & 2.33  \\
$B_{2g}$  &  770  &  1.79  & 5.30    & 8.88   & 3.58   & -5.67   & -4.69   & 0.97  \\
$B_{1g}$  &  796  &  1.73  & 5.47    & 8.29   & 2.82   & -5.69   &  -5.20  & 0.48  \\
\hline
\end{tabular}
\label{table:szo_gamma_delta3_raman}
\end{table*}

\begin{table*}[htp]
\centering
\caption{SZO data of the calculated IR frequencies at $T = 0 $ K, mode Gr\"uneisen parameter ($\gamma_j$),  linewidths ($\Gamma_j^{(3)}(T)$) and lineshifts ($\Delta_j^{(3)}(T)$) calculated at $T = 0$ and $T = 300$ K and their differences, i.e. $d\Gamma^{(3)}_j = \Gamma_j^{(3)}(300 \text{K}) - \Gamma_j^{(3)}(0 \text{K})$ and $d\Delta^{(3)}_j = \Delta_j^{(3)}(300 \text{K}) - \Delta_j^{(3)}(0 \text{K})$. Frequencies, linewidths, and lineshifts are given in units of cm$^{-1}$.}
\begin{tabular}{c c c c c c c c c}
\hline
Mode & $\omega$ & $\gamma_j$ &$\Gamma_j^{(3)}$ (0 K) & $\Gamma_j^{(3)}$ (300 K)  & $d\Gamma^{(3)}_j$  & $\Delta_j^{(3)}$ (0 K)  & $\Delta_j^{(3)}$ (300 K)  &  $d\Delta^{(3)}_j$ \\
\hline
$B_{1u}$   & 100  &  1.92  & 0.01    &  1.19   & 1.18   & -0.70   & -3.71   & -3.00   \\
$B_{3u}$   & 106  &  1.82  &  0.03   & 2.44   & 2.41   &  -1.41  & -5.84   & -4.43  \\
$B_{2u}$   & 113  &  2.11  & 0.30    & 4.90   & 4.60   & -1.50   & -5.52   & -4.02  \\
$B_{2u}$   & 137  &  1.09  &  0.44   & 8.20   & 7.76   & -2.92   & -10.75   & -7.83   \\
$B_{3u}$   & 153  &  2.15  &  0.30   & 5.40   & 5.09   & -2.01   & -7.31   & -5.31  \\
$B_{1u}$   & 155  &  0.64  & 0.45    & 7.05   & 6.60   & -2.80   & -9.05   & -6.25  \\
$B_{3u}$   & 193  &  0.30  &  0.62   & 6.11   &  5.49  & -1.79   & -4.32   & -2.54  \\
$B_{1u}$   & 204  &  0.92  & 0.68    & 5.32   & 4.64   & -2.27   & -6.30   & -4.03  \\
$B_{3u}$   & 209  & -0.33  & 0.75    & 5.87   & 5.12   & -2.10   & -5.19   & -3.09  \\
$B_{2u}$   & 211  &  0.46  &  0.75   & 6.95   & 6.21   & -2.04   & -6.34   & -4.30  \\
$B_{1u}$   & 213  & -0.17  & 0.29    & 2.62   & 2.33   & -1.47   & -4.83   & -3.36  \\
$B_{2u}$   & 244  &  0.88  &  0.80   & 5.50   & 4.71   & -2.34   & -5.33   & -2.98  \\
$B_{3u}$   & 256  &  1.31  &  0.60   & 4.47   & 3.87   & -2.14   & -5.07   & -2.93  \\
$B_{1u}$   & 260  &  0.48  &  0.64   & 4.61   & 3.97   & -1.92   & -4.26   & -2.34  \\
$B_{3u}$   & 279  &  1.66  & 0.55    & 3.28   & 2.73   & -1.06   & -2.22   & -1.16  \\
$B_{1u}$   & 329  &  0.89  &  0.48   & 4.12   & 3.64   & -2.60   & -4.46   & -1.86  \\
$B_{2u}$   & 333  &  1.73  & 0.66    & 4.02   & 3.36   & -2.64   & -4.75   & -2.12  \\
$B_{1u}$   & 339  &  1.09  & 1.12    & 4.85   & 3.73   & -2.92   & -4.88   & -1.96  \\
$B_{1u}$   & 364  &  0.55  & 1.51    & 5.36   & 3.85   & -1.80   & -2.92   & -1.12  \\
$B_{3u}$   & 386  &  0.65  & 1.02    & 3.62   & 2.60   & -1.10   & -1.34   & -0.24  \\
$B_{3u}$   & 462  &  0.48  & 1.48    & 6.65   & 5.17   & -2.20   & -3.61   & -1.40  \\
$B_{2u}$   & 492  &  3.22  & 1.04    & 5.69   & 4.65   & -3.32   & -4.42   & -1.10  \\
$B_{3u}$   & 505  &  2.73  & 0.91    & 4.52   & 3.61   & -3.31   & -4.24   & -0.93  \\
$B_{2u}$   & 512  &  2.66  & 0.98    & 4.59   & 3.61   & -3.16   & -3.34   & -0.18  \\
$B_{1u}$   & 521  &  2.55  & 1.12    & 3.71   & 2.59   & -2.93   & -3.65   & -0.73  \\
\hline
\end{tabular}
\label{table:szo_gamma_delta3_ir}
\end{table*}

\begin{table*}[htp]
\centering
\caption{SZO data of the calculated frequencies of inactive modes at $T = 0 $ K, mode Gr\"uneisen parameter ($\gamma_j$),  linewidths ($\Gamma_j^{(3)}(T)$) and lineshifts ($\Delta_j^{(3)}(T)$) calculated at $T = 0$ and $T = 300$ K and their differences, i.e. $d\Gamma^{(3)}_j = \Gamma_j^{(3)}(300 \text{K}) - \Gamma_j^{(3)}(0 \text{K})$ and $d\Delta^{(3)}_j = \Delta_j^{(3)}(300 \text{K}) - \Delta_j^{(3)}(0 \text{K})$. Frequencies, linewidths, and lineshifts are given in units of cm$^{-1}$.}
\begin{tabular}{c c c c c c c c c}
\hline
Mode & $\omega$ & $\gamma_j$ &$\Gamma_j^{(3)}$ (0 K) & $\Gamma_j^{(3)}$ (300 K)  & $d\Gamma^{(3)}_j$  & $\Delta_j^{(3)}$ (0 K)  & $\Delta_j^{(3)}$ (300 K)  &  $d\Delta^{(3)}_j$ \\
\hline
$A_{u}$   &  98   & 0.98   &  1.03   & 29.74   & 28.71   &  -5.51  &  -31.37  & -25.86   \\
$A_{u}$   &  131  & 1.09   & 0.04    & 1.63   & 1.59   & -1.07   & -4.10   & -3.03  \\
$A_{u}$   &  148  & 0.46   &  0.35   & 3.66   & 3.32   & -1.89   & -6.29   & -4.40  \\
$A_{u}$   &  195  & -0.29  &  0.64   & 7.09   & 6.46   & -1.62   & -5.31   & -3.69  \\
$A_{u}$   &  245  &  0.33  & 1.57    & 8.00   & 6.43   & -2.80   & -5.45   & -2.64  \\
$A_{u}$   &  328  &  1.24  & 0.59    & 3.63   & 3.04   & -2.76   & -5.49   & -2.73  \\
$A_{u}$   &  487  &  3.46  & 0.92    & 9.59   & 8.67   & -4.04   & -5.46   & -1.42  \\
$A_{u}$   &  509  &  2.84  & 0.87    & 4.47   & 3.60   & -3.58   &  -4.15  & -0.58  \\
\hline
\end{tabular}
\label{table:szo_gamma_delta3_silent}
\end{table*}

\begin{figure}[h]
 \centering
 \includegraphics[width=0.45\textwidth]{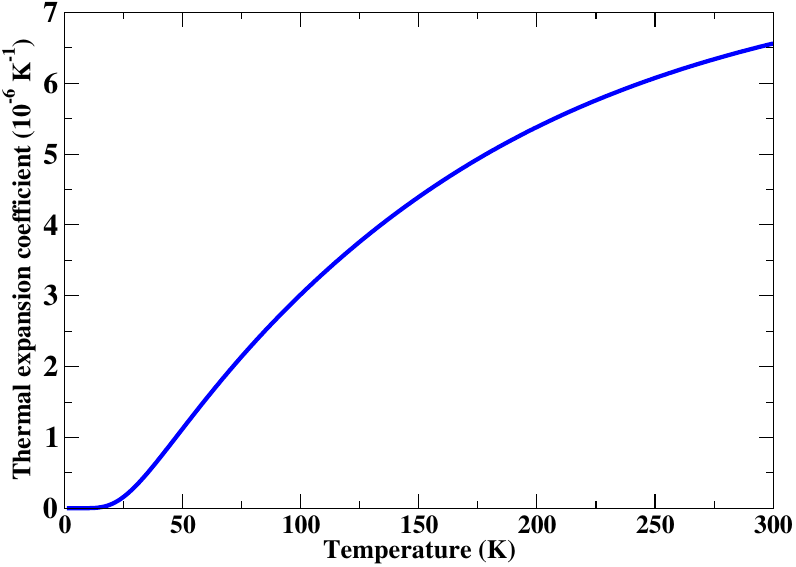}
 \caption{(Color online) Temperature dependence of the linear thermal expansion coefficient of orthorhombic SZO.}
 \label{fig:szo_alpha_vs_temp}
 \end{figure}

\begin{figure}[h]
  \centering
  \includegraphics[width=0.45\textwidth]{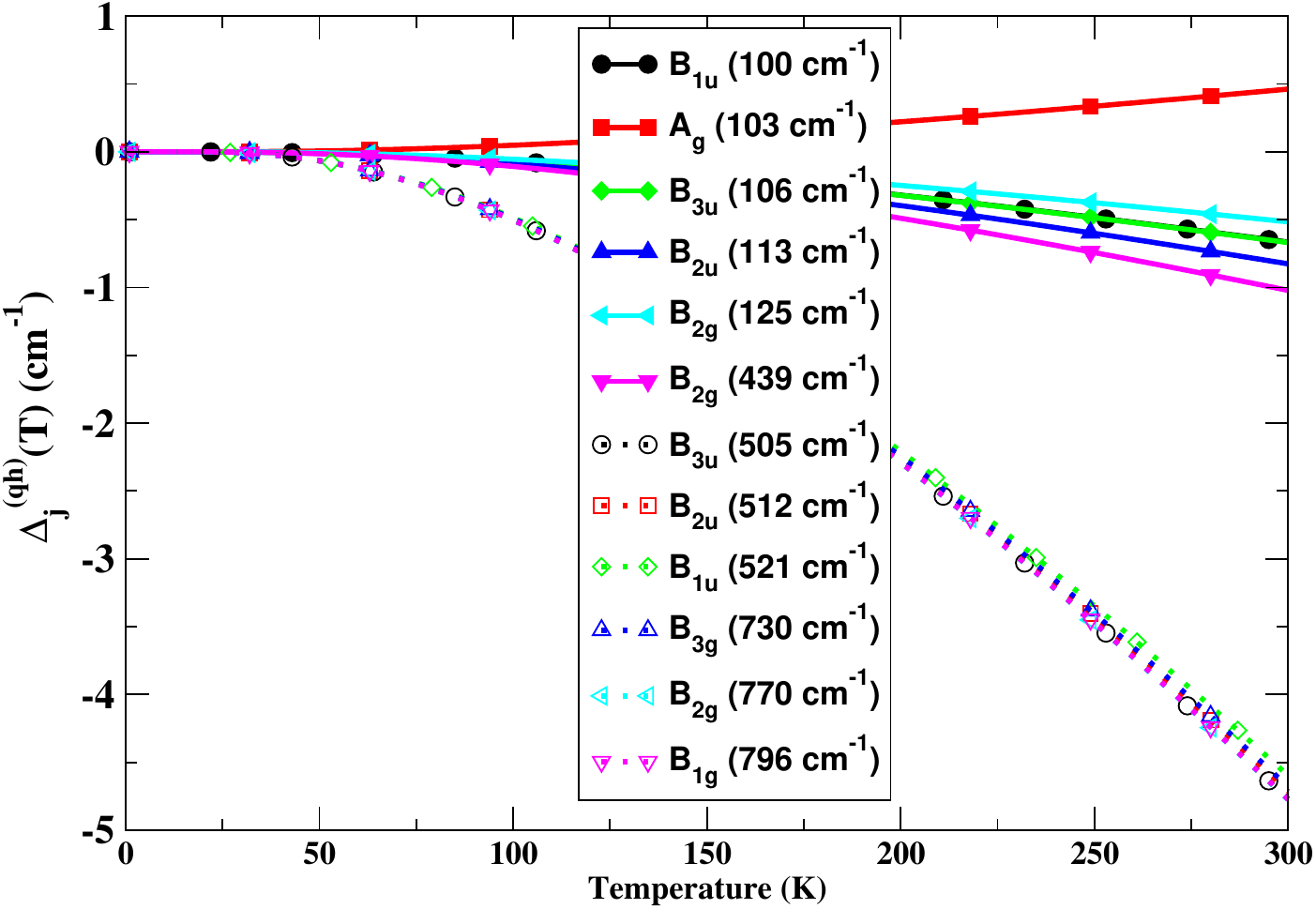}
  \caption{(Color online) Temperature dependence of the quasiharmonic shifts of optical modes in SZO. }
  \label{fig:szo_delta2}
\end{figure}

Next, we calculate the temperature dependence of the linear thermal expansion coefficient $\alpha(T)$ using Eq.~\ref{eqn:alpha}. It requires the evaluation of the isothermal bulk modulus $B$, which was already calculated and discussed in section~\ref{subsec:structural}. Furthermore, the computation of $\alpha(T)$ also requires the wavevector dependence of the mode Gr\"uneisen parameters $\gamma_j(\bm q)$, which in turn were calculated by performing the phonon dispersion calculations at several values of cell volume and fitting the frequencies of each mode using the cubic polynomials and consequently calculating the derivatives of $\omega_j(\bm q)$ with respect to the cell volumes. Thus the temperature dependence of the linear thermal expansion coefficient of orthorhombic SZO is shown in Fig.~\ref{fig:szo_alpha_vs_temp}. The thermal expansion coefficient is temperature-independent below 25 K, increases with increasing temperature, and finally, it is expected to saturate at higher temperatures. However, as noted earlier, we have limited our study to the temperature range of 0-300 K. Our calculated linear thermal expansion coefficient $\alpha(300\text{K}) \sim 0.66\times 10^{-5} \text{K}^{-1}$ at 300 K is reasonably close to the other calculated value of $\sim 0.76\times 10^{-5} \text{K}^{-1}$ \cite{KNIGHT201690}. Furthermore, the temperature dependence of the quasiharmonic shifts in frequencies $\Delta_j^{(qh)}(T)$ for a total of twelve optical modes including some of the low- and high-frequency modes ($B_{1u} (100~\text{cm}^{-1})$, $A_{g} (103~\text{cm}^{-1})$, $B_{3u} (106~\text{cm}^{-1})$, $B_{2u} (113~\text{cm}^{-1})$, $B_{2g} (125~\text{cm}^{-1})$, $B_{2g} (439~\text{cm}^{-1})$, $B_{3u} (505~\text{cm}^{-1})$, $B_{2u} (512~\text{cm}^{-1})$, $B_{1u} (521~\text{cm}^{-1})$, $B_{3g} (730~\text{cm}^{-1})$, $B_{2g} (770~\text{cm}^{-1})$, $B_{1g} (796~\text{cm}^{-1}))$ were computed using Eq.~\ref{eqn:qh}. Except for the one low-frequency mode $A_{g} (103~\text{cm}^{-1})$, the shifts $\Delta_j^{(qh)}(T)$ show a downward trend with increasing temperature. The mode $A_{g} (103~\text{cm}^{-1})$ behaves interestingly with temperature, where the $\Delta_j^{(qh)}(T)$ increases with temperature. This is due to its negative value of the mode Gr\"uneisen parameter. Next, $\Delta_j^{(qh)}(T)$ shows a large change for the high-frequency modes compared to the low-frequency modes, arising due to the large values of the mode Gr\"uneisen parameter of these modes. We notice that the shift $\Delta_j^{(qh)}(T)$ is negligibly small for all the modes in the temperature range of 0-25 K, which arises due to the vanishingly small value of the thermal expansion coefficient in this temperature range. 

\begin{figure}[htp]
  \centering
  \includegraphics[width=0.45\textwidth]{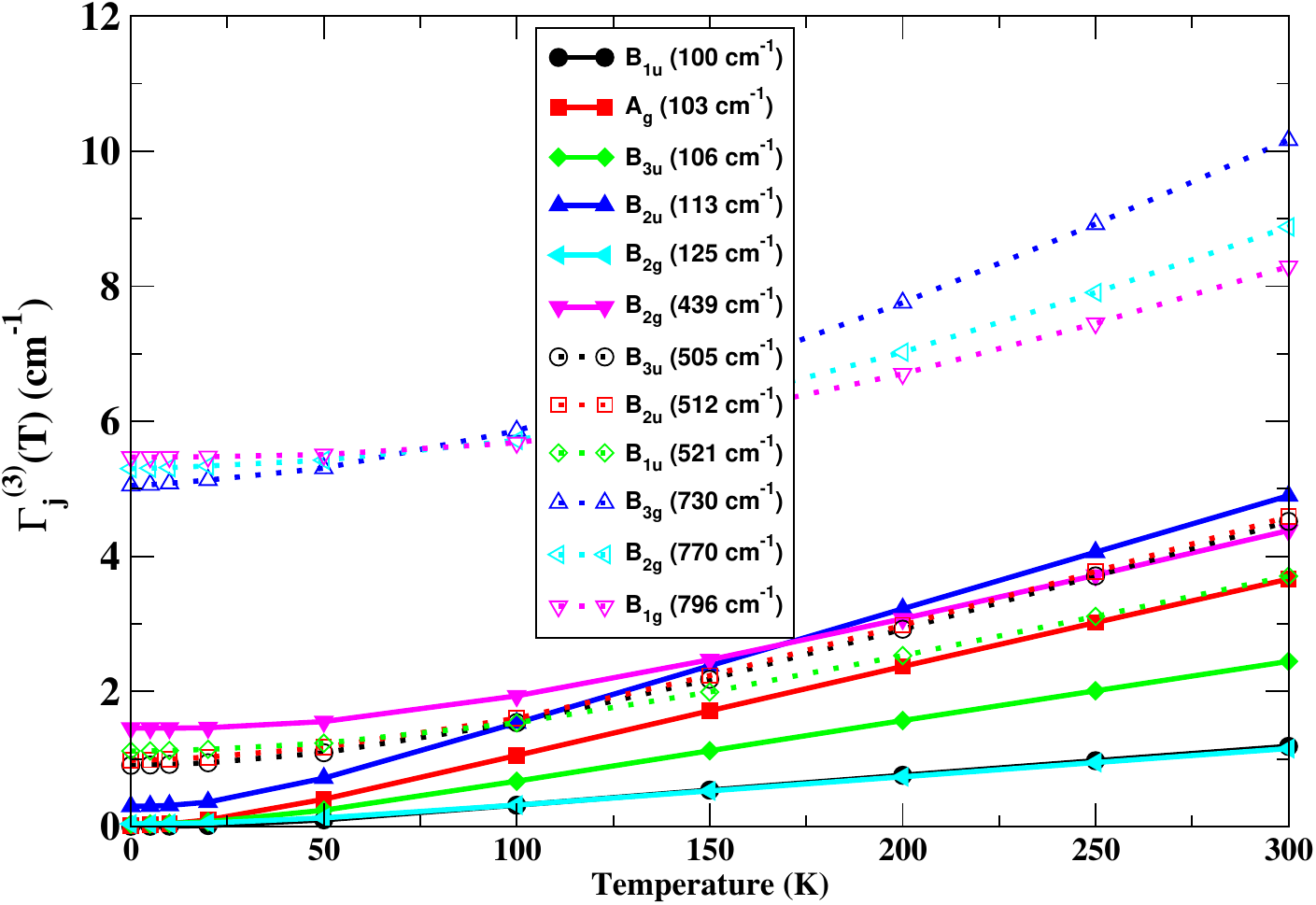}
  \caption{(Color online) Temperature dependence of the linewidths of optical modes in SZO. }
  \label{fig:szo_linewidth}
\end{figure}

\begin{figure}[htp]
  \centering
  \includegraphics[width=0.45\textwidth]{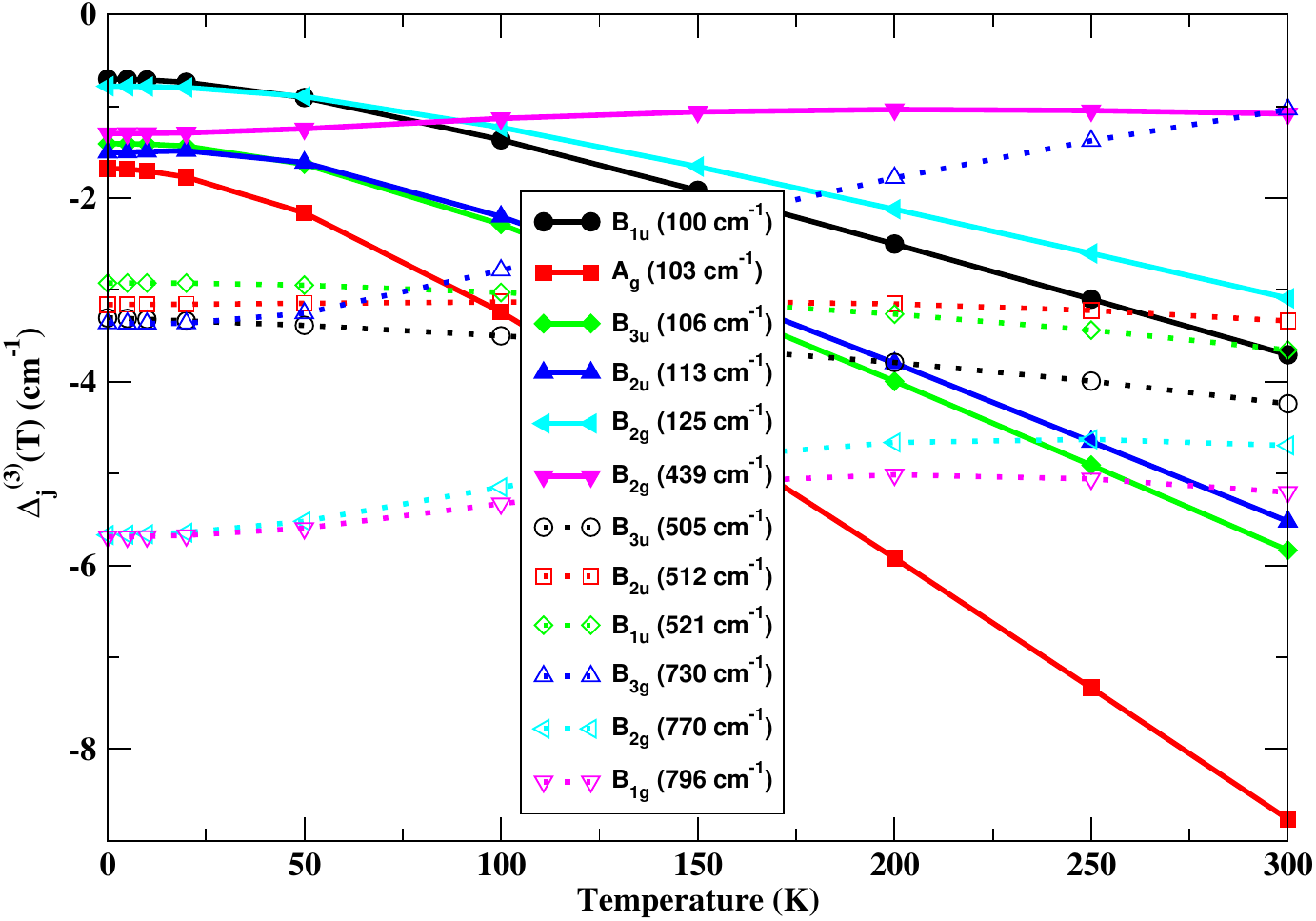}
  \caption{(Color online) Temperature dependence of the cubic anharmonic shifts of optical modes in SZO. }
  \label{fig:szo_delta3}
\end{figure}

\begin{figure}[htp]
  \centering
  \includegraphics[width=0.5\textwidth]{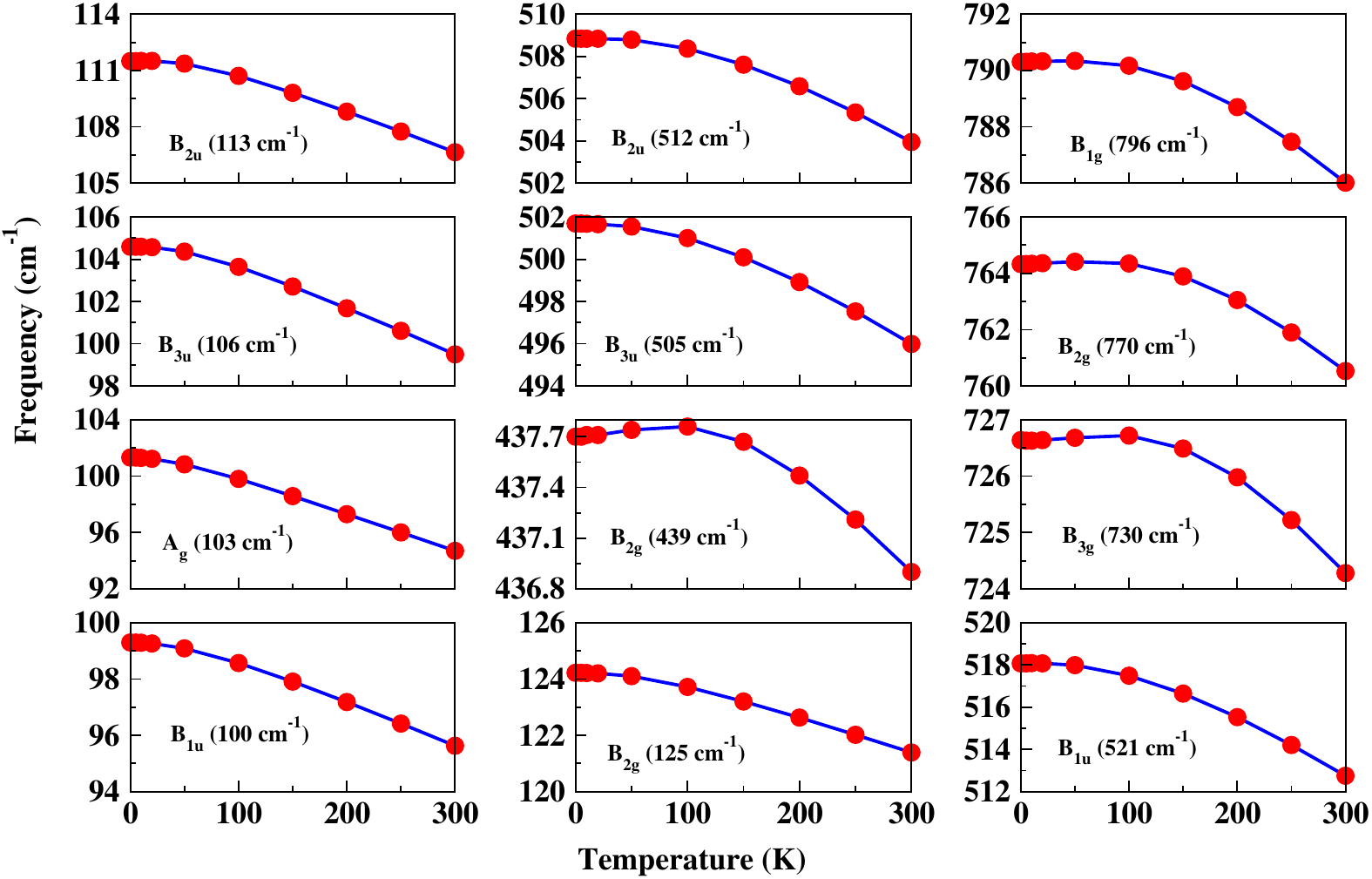}
  \caption{(Color online) Temperature dependence of frequencies of optical modes in SZO. }
  \label{fig:szo_freq}
\end{figure}

Next, the temperature dependence of the linewidths $\Gamma^{(3)}_j(T)$ of the twelve optical modes listed above is shown in Fig.~\ref{fig:szo_linewidth}. As can be seen from Eq.~\ref{eqn:Gamma}, $\Gamma^{(3)}_j(T)$ is temperature independent at lower temperatures and it goes linearly with temperature $\Gamma_j^{(3)}(T) \propto T$ at higher temperatures. Thus, $\Gamma^{(3)}_j(T)$ is temperature independent in the temperature range of 0-25 K. Furthermore, the temperature dependence of the lineshifts $\Delta^{(3)}_j(T)$ is shown in Fig.~\ref{fig:szo_delta3}. All modes exhibit non-zero values of $\Delta_j^{(3)}$(0 K) at zero Kelvin, which arise due to quantum corrections in energies at lower temperatures. For most modes, lineshifts show a downward trend with increasing temperature, except for a few high-frequency modes, such as $B_{2g}$ (439 cm$^{-1}$), $B_{3g}$ (730 cm$^{-1}$), $B_{2g}$ (770 cm$^{-1}$), and $B_{1g}$ (796 cm$^{-1}$). Although these modes initially exhibit upward shifts in $\Delta^{(3)}_j(T)$, which could in principle lead to anomalous behavior in their frequencies with temperature (softening with cooling), the contributions from all perturbative corrections to the harmonic level ultimately lead to a downward trend in frequency with increasing temperature for all modes in SZO, as shown in Fig.~\ref{fig:szo_freq}.  To the best of our understanding, there is no literature data on the temperature dependence of frequencies in orthorhombic SrZrO$_3$ perovskite in the temperature range of 0-300 K to compare with our results. However, there exists an experimental study \cite{AshokNPJCM2017} in this system on the temperature-dependent phonon frequencies, but in the temperature range of 300-970 K and also only for a few low-frequency modes. There, the authors have also observed a similar trend of the normal temperature dependence of the frequencies with temperature, i.e. hardening with cooling. We note that the SZO remains in the orthorhombic phase until 970 K.

\section{Two Phonon DOS and Kinematical Function}
\label{sec:tphdos_kf}

Assuming the independence of the three-phonon matrix elements in Eq.~\ref{eqn:Gamma} with respect to its arguments, the linewidth becomes proportional to the two-phonon density of states (TDOS), denoted as $D(\omega, T)$~\cite{PKVermaPRB2022}.

\begin{eqnarray}
D(\omega, T) &=& \frac{1}{N}\sum_{\bm{q},j_1,j_2} \bigg\{[n(\omega_{j_1}(\bm{q}))+ n(\omega_{j_2}(\bm {-q})) + 1]\times  \nonumber \\
 & & [ \delta(\omega - \omega_{j_1}(\bm{q}) - \omega_{j_2}(\bm{-q})) \nonumber \\
 && - \delta(\omega + \omega_{j_1}(\bm{q}) + \omega_{j_2}(\bm{-q})) ] \nonumber \\
& & + 2 [n(\omega_{j_2}(\bm{-q})) -  n(\omega_{j_1}(\bm {q}))]\times \nonumber \\
& & \delta(\omega - \omega_{j_1}(\bm{q}) + \omega_{j_2}(\bm{-q})) \bigg \}
\label{eqn:tphdos}
\end{eqnarray}

The first and second set of terms in the curly bracket is commonly known as down-conversion and up-conversion phonon processes, respectively~\cite{PKVermaPRB2022}. 
     
Under the same approximation, the line-shift functions are also independent of $j$ and are proportional to the two-phonon kinematical function denoted as $P(\omega, T)$. This function corresponds to the Hilbert transform of $D(\omega, T)$ and is expressed as:

\begin{eqnarray}
P(\omega, T) &=& \frac{1}{N}\sum_{\bm{q},j_1,j_2}  \mathcal{P} \Bigg\{ \frac{n(\omega_{j_1}(\bm{q}))+ n(\omega_{j_2}(\bm {-q})) + 1} {\omega - \omega_{j_1}(\bm{q}) - \omega_{j_2}(\bm{-q})} \nonumber \\
 & &   - \frac{n(\omega_{j_1}(\bm{q}))+ n(\omega_{j_2}(\bm {-q})) + 1} {\omega + \omega_{j_1}(\bm{q}) + \omega_{j_2}(\bm{-q})}  \nonumber \\
& & + 2 \frac{ n(\omega_{j_2}(\bm{-q})) -  n(\omega_{j_1}(\bm {q}))} { \omega - \omega_{j_1}(\bm{q}) + \omega_{j_2}(\bm{-q})}  \Bigg \}
\label{eqn:kinematic}
\end{eqnarray}
where $\mathcal{P}$ represents the Cauchy principal part. As mentioned earlier, the temperature dependence of $D(\omega, T)$ and $P(\omega, T)$ originates from the phonon occupancy factors $n$.

\begin{figure}[htp]
  \centering
  \includegraphics[width=0.45\textwidth]{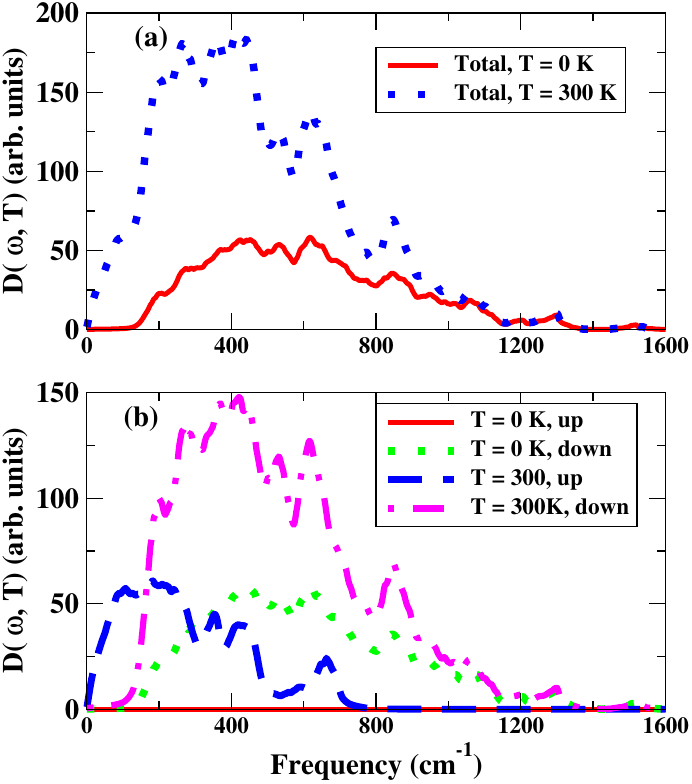}
  \caption{(Color online) Two phonon density of states of orthorhombic SrZrO$_3$ computed at 0 and 300 K of temperatures. (a) Total contribution and (b) the contributions arising from  the up and down conversion decay processes as mentioned in the main text. }
  \label{fig:szo_tphdos}
\end{figure}

Fig.~\ref{fig:szo_tphdos} displays the TDOS calculated at two temperatures, $T = 0$ and $300$ K. Additionally, to gain further insights, we have plotted the separate contributions from up-conversion and down-conversion processes in $D(\omega, T)$ for the same temperatures. At high frequencies, the up-conversion contributions are notably smaller compared to down-conversion due to fewer available channels for the former. In contrast, the down-conversion process exhibits a wide frequency window with a significantly larger TDOS, particularly evident at all temperatures.

\begin{figure}[htp]
  \centering
  \includegraphics[width=0.45\textwidth]{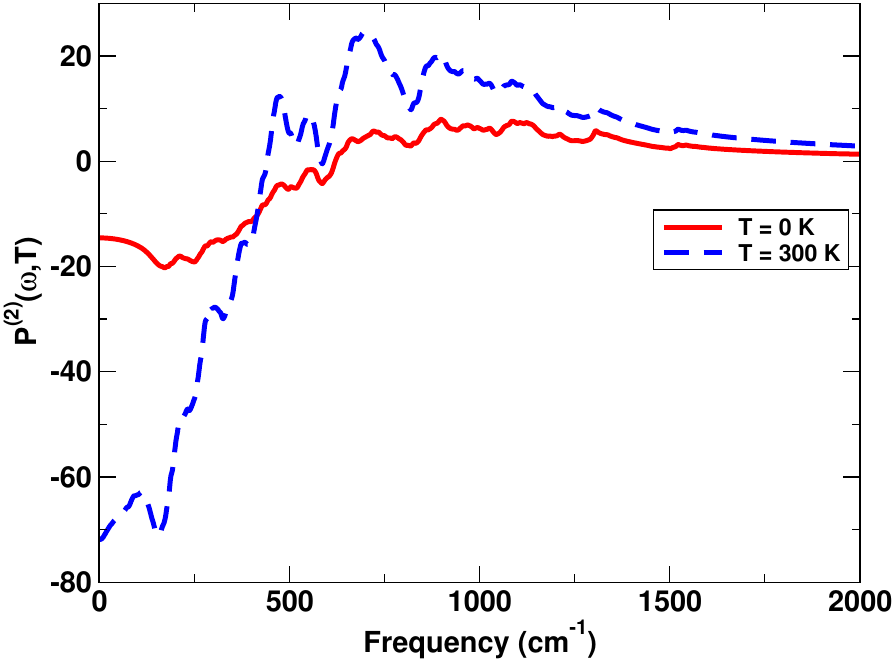}
  \caption{(Color online) Frequency dependence of the two phonon kinematical function $P(\omega,T)$ of orthorhombic SrZrO$_3$ calculated at two different values of temperatures 0 and 300 K.}
  \label{fig:szo_kinematic}
\end{figure}

Next, Fig.~\ref{fig:szo_kinematic} illustrates the frequency dependence of the two-phonon kinematical function $P(\omega, T)$ calculated at two distinct temperatures. The zero Kelvin value of $P(\omega, T)$ is almost negative for all the modes present in the system. However, it shows a positive value of some of the high-frequency modes at 300 K of temperature. Though, $P(\omega, T)$ gives a qualitative picture of the shift of phonon frequencies, it has no one-to-one correspondence with the shifts $\Delta^{(3)}(\omega, T)$ as the coupling constant matrix elements $V^{(3)}(\bm 0, j;\bm q, j_1;-\bm q, j_2)$ play a significant role in determining the actual temperature-dependent behavior of the frequency shifts.
   
\section{Conclusions}
\label{sec:conclusions}

This paper presented a comprehensive study of the temperature-dependent phonon properties in orthorhombic SZO perovskite using first-principles density functional theory calculations. Our study includes a perturbative approach to investigate the effects of lattice anharmonicity on phonon properties. We found that the frequencies of the modes were consistent with previous theoretical data, but discrepancies were observed in the frequencies reported by Vali et al. \cite{VALI2008497}. Our calculated Raman active mode frequencies were in good agreement with available experimental results, although there was no experimental data on the IR frequencies to compare with our results.

Furthermore, we discussed the G\"uneisen parameters, linear thermal expansion coefficient, and frequency shifts of the modes within the quasiharmonic approximation. We found that the modes exhibit a downward shift with increasing temperature. We also presented the temperature dependence of phonon linewidths and lineshifts within the third-order lattice anharmonic effect. We found that the lineshifts of some modes show an upward trend with increasing temperature, which could lead to the anomalous behavior of the frequencies of these modes with temperature (softening with cooling). However, when both the quasiharmonic and third-order anharmonic shifts were added to the zeroth contribution within the harmonic approximation, almost all the modes behaved normally with temperature, i.e. hardening of frequencies with cooling.

\section*{ACKNOWLEDGMENTS}
The author would like to acknowledge the Council of Scientific and Industrial Research, India for the financial support provided for this research. The author would also like to acknowledge the Supercomputer Education and Research Centre at IISc for providing the necessary computational resources for the calculations performed on the Cray XC40 system.

\newpage

 \bibliography{bibl}  

\end{document}